\documentclass{elsart}

\usepackage{natbib}

\usepackage{graphicx}

\usepackage{epsfig}

\usepackage{amssymb}

\usepackage{lineno}

\begin{document}

\font\ksmss=cmss10     scaled \magstephalf
\font\kvsmss=cmss8     scaled \magstephalf

\newcommand{\cge}    {{$_ >\atop{^\sim}$}}
\newcommand{\cle}    {{$_ <\atop{^\sim}$}}
\newcommand{\n}	     {\noindent}%
\newcommand{\si}	 {\smallskip\indent}%
\newcommand{\sn}	 {\smallskip\noindent}%
\newcommand{\mi}	 {\medskip\indent}%
\newcommand{\mn}	 {\medskip\noindent}%
\newcommand{\bi}	 {\bigskip\indent}%
\newcommand{\bn}	 {\bigskip\noindent}%
\newcommand{\cl}     {\centerline}%
\newcommand{\ve}     {\vfill\eject}
\newcommand{\eg}	 {\mbox{e.g.,} }%
\newcommand{\cf}	 {\mbox{c.f.,} }%
\newcommand{\ie}	 {\mbox{i.e.} }%
\newcommand{\aka}	 {\mbox{a.k.a.\ }}%
\newcommand{\bul}    {$\bullet$\ }
\newcommand\Bj       {{$B_{J}$}}
\newcommand{\HST}	 {\emph{HST}}%
\newcommand{\Ha}	 {\emph{H-$\alpha$}}%
\newcommand\JAB      {{$J_{AB}$}}
\newcommand\Lstar    {{$L^{*}$}}
\newcommand\mum      {{$\mu$m}}
\newcommand\Msun     {{\ $M_{\odot}$}}
\newcommand\zform    {{$z_{form}$}}
\newcommand\zAB      {{$z_{AB}$} }
\newcommand{\Astro}	 {\emph{Astro}}%
\newcommand\sspt     {{$\buildrel{s}           \over .$}}
\newcommand\degpt    {{$\buildrel{\circ}       \over .$}}
\newcommand\adegpt   {{$\buildrel{\circ}       \over .$}}
\newcommand\arcmpt   {{$\buildrel{\prime}      \over .$}}
\newcommand\arcspt   {{$\buildrel{\prime\prime}\over .$}}
\newcommand\magpt    {{$\buildrel{m}           \over .$}}
\newcommand\adeg     {{\hskip .10em {}^{\circ}\hskip -.40em . \hskip.25em} }
\newcommand\amin     {{\hskip .10em {}'       \hskip -.30em . \hskip.20em} }
\newcommand\asec     {{\hskip .10em {}''      \hskip -.49em . \hskip.29em} }
\newcommand\A        {{{\kvsmss A}{\hspace*{-2.0mm}\vspace*{-0.0mm}$^{\scriptsize o}$}
\vspace*{+0.0mm}}}
\newcommand\degsq    {{\ $deg^{2}$} }
\newcommand{\kmsMpc}	{\mbox{km s$^{-1}$ Mpc$^{-1}$}}%
\newcommand{\magarc}	{{$\mbox{mag arcsec}^{-2}$}}%
\newcommand{\SB}	 {{$\mu_{\scriptsize\sl F300W}$}}%
\newcommand{\re}	 {{$r_e$}}%
\newcommand{\rhl}	 {{$r_{hl}$} }%
\newcommand{\HI}	 {\mbox{H\,{\sc i}}}%
\newcommand{\HII}	 {\mbox{H\,{\sc ii}}}%
\newcommand\Kmorph   {{$K_{morph}$} }
\newcommand{\UIT}	 {\emph{UIT}}%
\newcommand\zmed     {{$z_{med}$}}
\newcommand{\tsim}	 {{\sim}}%
\newcommand{\lsim}	 {\makebox[15pt]
           {\mbox{{\raisebox{-3pt}{$\sim$}}{\llap{\raisebox{2pt}{$<$}}}}}}%
\newcommand{\gsim}	 {\makebox[15pt]
           {\mbox{{\raisebox{-3pt}{$\sim$}}{\llap{\raisebox{2pt}{$>$}}}}}}%
\newcommand\degree   {{\ifmmode^\circ\else$^\circ$\fi}} 

\newcommand\AAP      {A\&A, }
\newcommand\aap      {A\&A}
\newcommand\AAL      {A\&AL, }
\newcommand\aal      {A\&AL}
\newcommand\AAPS     {A\&ApS, }
\newcommand\aaps     {A\&ApS}
\newcommand\AAS      {A\&AS, }
\newcommand\aas      {A\&AS}
\newcommand\ACTAA    {Acta~A, }
\newcommand\AJ       {AJ, }
\newcommand\aj       {AJ}
\newcommand\AN       {AN, }
\newcommand\APLET    {Ap. Let., }
\newcommand\APJ      {ApJ, }
\newcommand\apj      {ApJ}
\newcommand\APJL     {ApJL, } 
\newcommand\apjl     {ApJL} 
\newcommand\APL      {ApJ, }  
\newcommand\APJS     {ApJS, }
\newcommand\apjs     {ApJS}
\newcommand\APS      {ApJS, }
\newcommand\APSS     {Ap\&SS, }
\newcommand\apss     {Ap\&SS}
\newcommand\ARAA     {ARA\&A, }
\newcommand\araa     {ARA\&A}
\newcommand\AUSJP    {Australian J. Phys., }
\newcommand\AZH      {AZh, }
\newcommand\BAAS     {BAAS, }
\newcommand\baas     {BAAS}
\newcommand\FCP      {Fund. Cosmic Phys., }
\newcommand\HOA      {Highlights Astr., }
\newcommand\IAU      {IAU Symp., }
\newcommand\ITR      {Internal Technical Report of the Netherlands Foundation for 
                      Radio Astronomy, No.} 
\newcommand\JOSAMA   {J. Opt. Soc. Am. A, }
\newcommand\JRASC    {JRASC, }
\newcommand\MEMRAS   {MmRAS, }
\newcommand\MNRAS    {MNRAS, }
\newcommand\mnras    {MNRAS}
\newcommand\NAT      {{\it Nature}, }
\newcommand\nat      {{\it Nature}}
\newcommand\NPS      {Nat. Phys. Sc., }
\newcommand\PASA     {PASA, }
\newcommand\pasa     {PASA}
\newcommand\PASJ     {PASJ, }
\newcommand\pasj     {PASJ}
\newcommand\PASP     {PASP, }
\newcommand\pasp     {PASP}
\newcommand\PHD      {Ph.D. thesis, }
\newcommand\PHYSCR   {Physica Scripta, }
\newcommand\QJRAS    {QJRAS, }
\newcommand\REP      {Reports on Astronomy, IAU Transactions, }
\newcommand\REPPPHY  {Rep. Prog. Phys., }
\newcommand\REVMP    {Rev. Mod. Phys., }
\newcommand\SCAM     {Sc. Am., }
\newcommand\SCI      {Science, }
\newcommand\SOVAST   {Sov. Astr., }
\newcommand\SPIE     {SPIE, }
\newcommand\ST       {S\&T, }
\newcommand\VIS      {Vistas in Astronomy, }
\newcommand\VLA      {VLA Test Memorandum, No.}

\newlength{\txw}
\setlength{\txw}{\textwidth}
\newlength{\txh}
\setlength{\txh}{\textheight}



\begin{frontmatter}



\title{High resolution science with high redshift galaxies}

\author{R. A. Windhorst, N. P. Hathi, S. H. Cohen, R. A. Jansen }
\address[label1] {School of Earth \& Space Exploration, Arizona State University,
Box 871404, Tempe, AZ 85287-1404, USA;\ \ \ Email:\ Rogier.Windhorst@asu.edu}

\author{D. Kawata}
\address[label2] {Carnegie Observatories, 813 Santa Barbara Street
Pasadena, CA 91101}

\author{S. P. Driver}
\address[label3] {School of Physics and Astronomy, 
St Andrews, Fife, KY16 9SS Scotland}

\author{B. Gibson}
\address[label4] {Univ. of Central Lancashire, 
Preston, Lancashire, PR1 2HE United Kingdom}


\begin{abstract} 

We summarize the high-resolution science that has been done on high redshift
galaxies with Adaptive Optics (AO) on the world's largest ground-based
facilities and with the Hubble Space Telescope (HST). These facilities
complement each other. Ground-based AO provides better light gathering power
and in principle better resolution than HST, giving it the edge in high
spatial resolution imaging and high resolution spectroscopy. HST produces
higher quality, more stable PSF's over larger field-of-view's in a much darker
sky-background than ground-based AO, and yields deeper wide-field images and
low-resolution spectra than the ground. Faint galaxies have steadily decreasing
sizes at fainter fluxes and higher redshifts, reflecting the hierarchical
formation of galaxies over cosmic time. HST has imaged this process in great
structural detail to z\cle 6, and ground-based AO and spectroscopy has provided
measurements of their masses and other physical properties with cosmic time.
Last, we review how the 6.5 meter James Webb Space Telescope (JWST) will
measure First Light, reionization, and galaxy assembly in the near--mid-IR
after 2013. 

\end{abstract}


\end{frontmatter}


\section{Introduction }

In this paper, we briefly review the current status of high resolution imaging
of high redshift galaxies. In the last decade, major progress has been made
with the Hubble Space Telescope (HST), and through targeted programs using
Adaptive Optics (AO) on the world's best ground-based facilities. It is not
possible to review all these efforts here, and so we refer the reader to more
detailed reviews in proceedings by, \eg Livio, Fall, \& Madau (1998), Cristiani,
Renzini, \& Williams (2001), Mather, MacEwen, \& de Graauw (2006), Ellerbroek
\& Bonaccini Calia (2006), and Gardner \etal\ (2006).

\section{What can and has been done from the ground? }

High resolution AO-imaging on distant galaxies has been carried out
successfully with large ground-based telescopes. A number of AO studies 
observed distant galaxies in the near-IR (\eg Larkin \etal\ 2000, 2006; Glassman \etal\
2002; Steinbring \etal\ 2004; Melbourne \etal\ 2005; and Huertas-Company \etal\
2007). Large ground-based telescopes with well calibrated AO {\it can} in
principle match or supersede HST's resolution on somewhat brighter objects than
accessible to HST, if AO guide stars are available in or nearby the AO
field-of-view (FOV), as shown by Steinbring \etal\ (2004; Fig. 1ab here).
Ground-based telescopes can also provide a much larger collecting area,
allowing one to obtain higher spectral resolution, spatially-resolved spectra
of faint galaxies (\eg Larkin \etal\ 2006). This enables the study of the
morphology and rotation curves of faint galaxies in order to measure their
masses and constrain galaxy assembly. Melbourne \etal\ (2005) used Keck AO and
HST images to distinguish stellar populations, AGN and dust (Fig. 1c here). At
longer wavelengths ($\lambda$\cge 1--2\mum), ground-based AO has provided PSF's
that are as good as, or sharper than the $\lambda$/D that the 2.4 meter HST
provides. 

The PSF-stability and dynamic range, FOV, low sky-brightness and depth that
diffraction limited space based images provide are difficult to match by
ground-based AO imaging. There are two primary factors for this. First,
atmospheric phase fluctuations (seeing) affect the Strehl ratio and
PSF-stability, and therefore the effective dynamic range and FOV of 
ground-based AO images, compared to the diffraction-limited PSF and FOV that
the (aberration corrected) HST provides. Second, the sky-brightness at
$\lambda$$\simeq$1--2\mum\ is typically $\sim$10$^{3}\times$ (or $\sim$7 mag)
fainter in space compared to the ground (Thompson \etal\ 2006). The bright
atmospheric OH-forest thus limits the surface brightness (SB) sensitivity that
can be achieved from the ground, even with larger telescopes. Without AO, the
deepest ground-based near-IR imaging achieved to date in the best natural
seeing ($\sim$0\arcspt 46 FWHM) was done with VLT/ISAAC

\noindent\leftline{
    \psfig{file=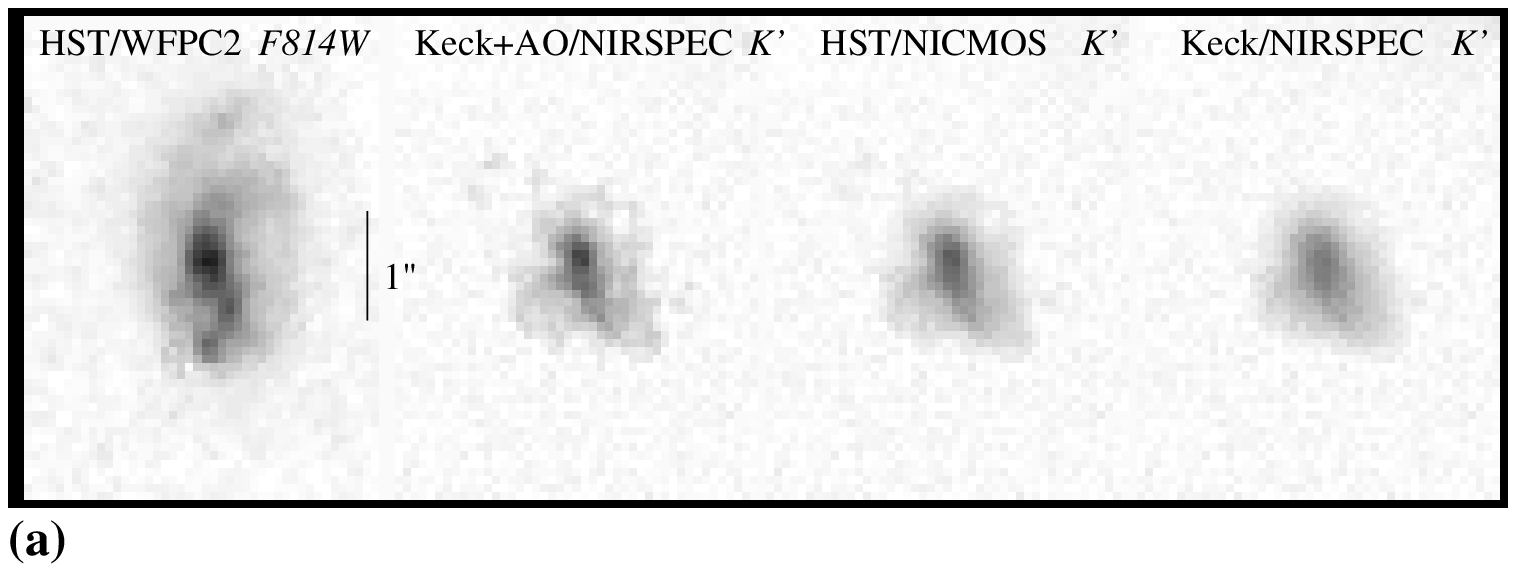,angle=0,width=0.57\txw}\ 
    \raisebox{-35pt}[0pt][-35pt]{
       \psfig{file=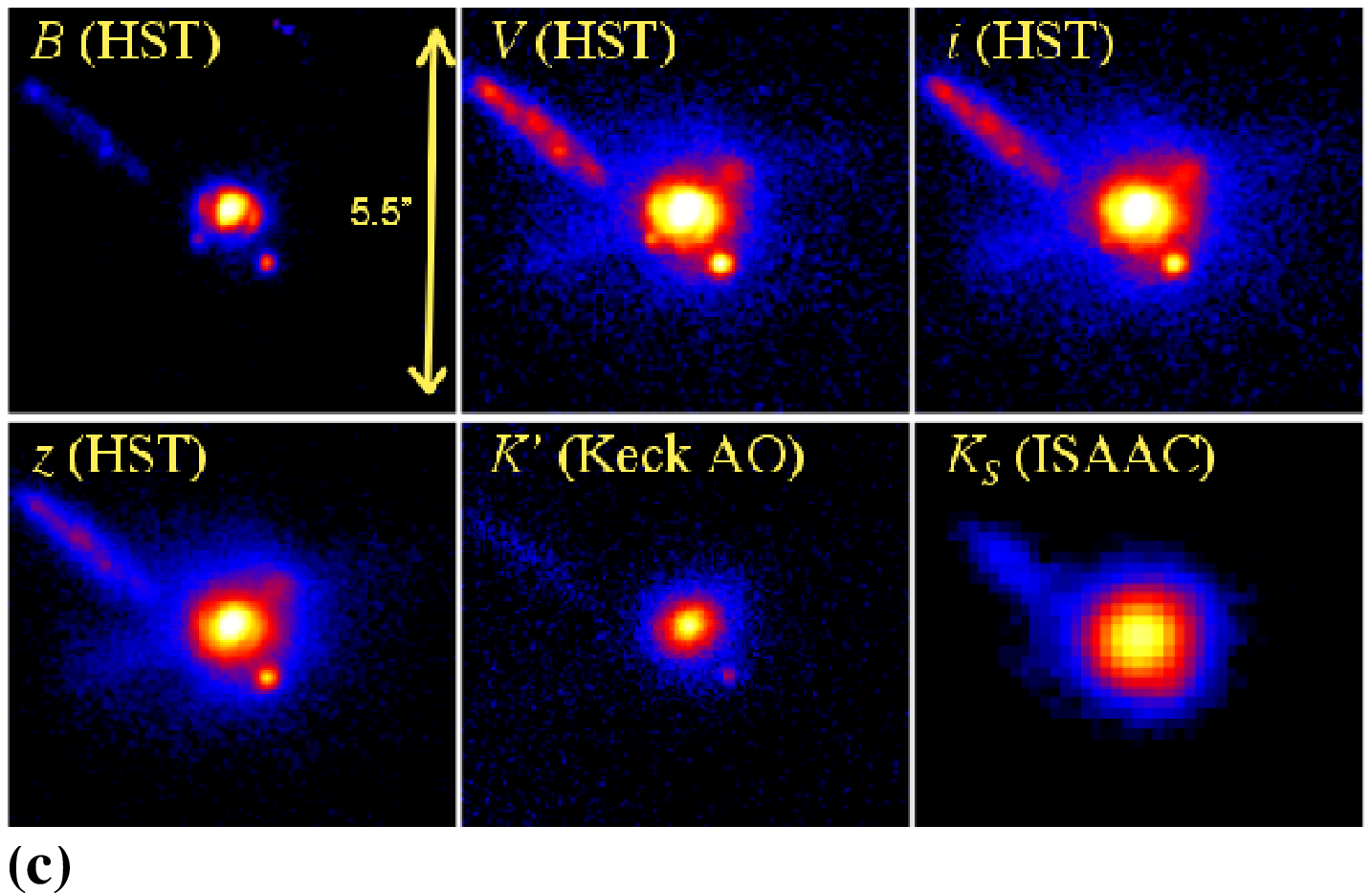,angle=0,width=0.40\txw,height=0.3\txw}
    }
}\\[2pt]
\leftline{
    \psfig{file=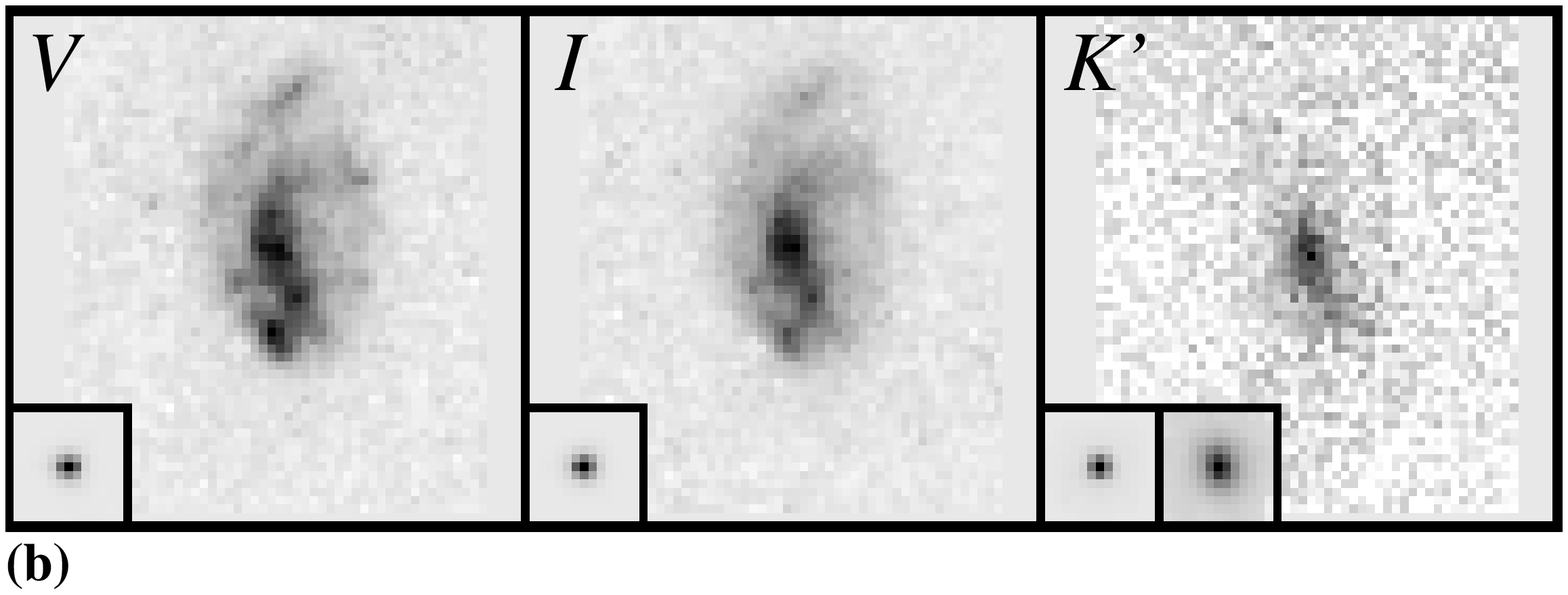,angle=0,width=0.57\txw}
}
\noindent{\footnotesize
{\bf Fig.~1ab} \ Comparison of Keck~AO images of a spiral galaxy at z=0.531
to HST V and I-band images, a simulated HST/NICMOS K$'$ image, and a Keck
NIRSPEC image in natural seeing (from Steinbring \etal\ 2004). 
{\bf Fig.~1c} \ Comparison of Keck~AO images of a recent merger at z=0.61
to HST and VLT/ISAAC images (from Melbourne \etal\ 2005).}\\

\noindent in the HDF-S (Labb\'e \etal\ 2003), reaching $J$=25.8, $H$=25.2 and
$K_s$=25.2 AB-mag (7.5\,$\sigma$) in $\sim$35 hours per filter. HST/NICMOS can
reach these sensitivities in less than one hour, or could reach \cge 2 mag
deeper in the same amount of time. These VLT images would have gone deeper, had
they been done with AO, but then they may not have covered a 2\arcmpt
5$\times$2\arcmpt 5 FOV. In conclusion, diffraction limited space-based imaging
provides much darker sky over a wider FOV, more stable PSF's, better dynamic
range, and therefore superior sensitivity. Ground-based AO is
\emph{complementary} to what space-based imaging can do. In the future,
multi-conjugate AO (MCAO) from the ground will aim to provide nearly
diffraction limited imaging over wider FOV's than possible with AO alone.
Hence, MCAO facilities on 8--30 meter telescopes may become competitive with
HST and JWST at 1--2 \mum\ wavelength in terms of PSF-width {\it and} FOV. This
is why JWST no longer has cost-driving specifications below 1.7 \mum\ 
wavelength, although it will probably perform quite well to 1.0 and possibly
0.7 \mum. Future MCAO may not be competitive with space-based imaging in terms
of PSF-stability, dynamic range, sky-brightness, and therefore sensitivity. In
the thermal infrared ($\lambda$\cge 2\mum), space-based imaging will be
superior in depth. But to achieve the highest possible resolution on somewhat
brighter objects, ground-based MCAO will be superior to space-based imaging.
{\it It is critical for the future development of both space-based and
ground-based high resolution imaging to keep this complementarity in mind, so
that both sets of instruments can be developed to maximize the overall
scientific return.} 

\ve

\noindent\begin{minipage}[b]{0.65\txw}
   \psfig{file=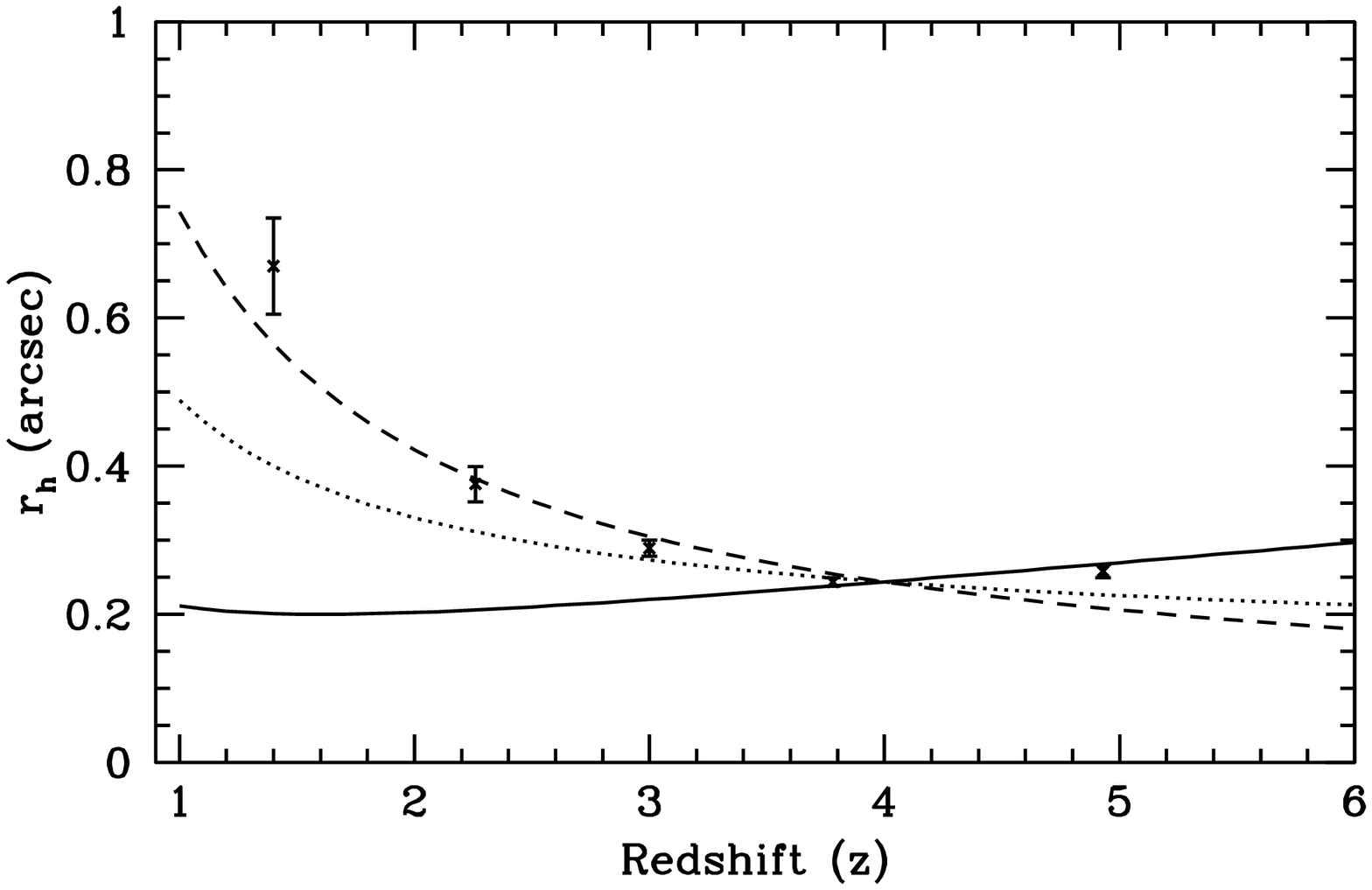, angle=0,width=0.625\txw}
\end{minipage}\hfill
\begin{minipage}[b]{0.32\txw}\footnotesize
{\bf Fig. 2} \ Size evolution of galaxies in the HST GOODS fields (from
Ferguson \etal\ 2004), indicated by the dashed and dotted curves, as
summarized in \S~3. The solid curve indicates constant sizes in WMAP
cosmology.
\end{minipage}

\hspace*{+0.0cm}\vspace*{-1.05cm}
\section{Why does high-resolution imaging need to be done from space?}

The HST/ACS GOODS survey (Ferguson \etal\ 2004) showed that the median sizes of
faint galaxies decline steadily towards higher redshifts (Fig. 2), despite the
$\Theta$--z relation that minimizes at z$\simeq$1.65 in WMAP $\Lambda$CDM
cosmology. While SB and other selection effects in these studies are
significant, this figure suggests evidence for intrinsic size evolution of
faint galaxies, where galaxy half-light radii \rhl evolve approximately with
redshift as:\ r$_{\rm hl}$(z)\,$\propto$\,r$_{hl}$(0)$\cdot$(1+z)$^{\rm -s}$\ 
with s\,$\simeq$\,1. This reflects the hierarchical formation of galaxies, 
where sub-galactic clumps and smaller galaxies merge over time to form the
larger/massive galaxies that we see today (\eg Navarro, Frenk, \& White 1996). 

The HST/ACS Hubble UltraDeep Field (HUDF; Beckwith \etal\ 2006) showed that
high redshift galaxies are intrinsically very small, with typical sizes of
$r_{\rm hl}$$\simeq$ 0\arcspt 12 or 0.7--0.9 kpc at z$\simeq$4--6. A
combination of ground-based and HST surveys shows that the apparent galaxy
sizes decline steadily from the RC3 to the HUDF limits (Fig. 3 here; Odewahn
\etal\ 1996; Cohen \etal\ 2003, Windhorst \etal\ 2006). At the bright end, this
is due to the survey SB-limits, which have a slope of +5 mag/dex in Fig. 3. At
the faint end, ironically, this appears {\it not} to be due to SB-selection
effects (cosmological (1+z)$^{4}$ SB-dimming), since for \Bj\cge 23 mag the
samples do {\it not} bunch up against the survey SB-limits. Instead it occurs
because: (a) their hierarchical formation and size evolution (Fig. 2); (b) at
\JAB\cge 26 mag, one samples the faint end of the luminosity function (LF) at
z$_{\rm med}$\cge 2--3, resulting in intrinsically smaller galaxies (Fig. 4b;
Yan \& Windhorst 2004b); and (c) the increasing inability to properly deblend
faint galaxies at fainter fluxes. This leads ultradeep surveys to slowly
approach the ``natural'' confusion limit, where a fraction of the objects
unavoidably overlaps with neighbors due to their finite \emph{object size}
(Fig. 3), rather than the finite instrumental resolution, which causes the
\emph{instrumental} confusion limit. Most

\ve



\noindent\begin{minipage}[b]{0.74\txw}
   \psfig{file=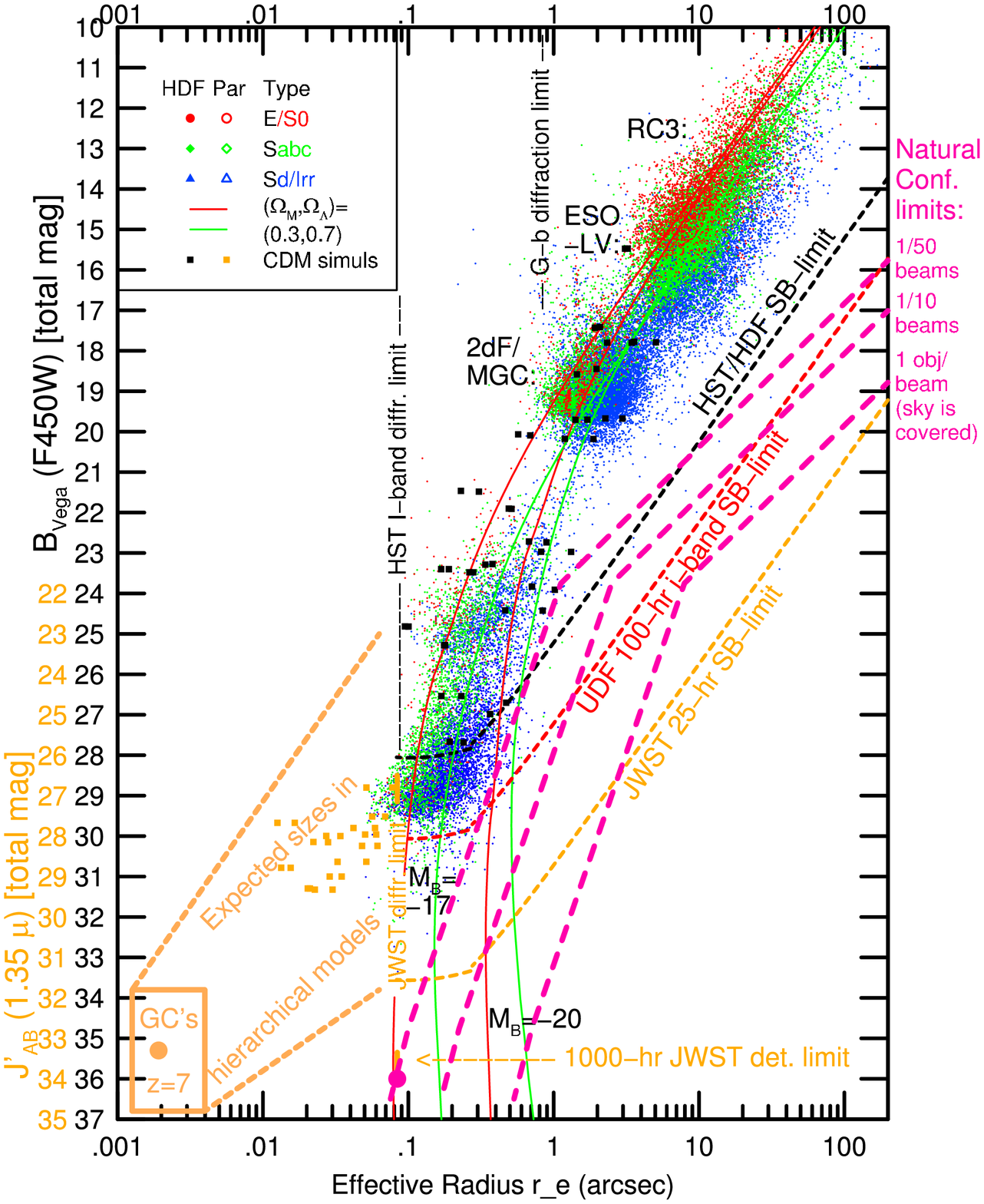,angle=0,width=0.73\txw,height=0.575\txh} 
\end{minipage}
\begin{minipage}[b]{0.2525\txw}
\noindent{\footnotesize
{\bf Fig. 3} \ Galaxy sizes vs. B$_{Vega}$ or \JAB-mag from the RC3 to the HUDF
limit. Short dashed lines indicate survey limits for the HDF (black), HUDF
(red), and JWST (orange): the point-source sensitivity is horizontal and the
SB-sensitivity has slope=+5 mag/dex. Broken long-dashed pink lines indicate the
natural confusion limit, below which objects begin to overlap due to their own
sizes. Red and green lines indicate the expectations at faint fluxes of the
{\it non-evolving} median size for RC3\linebreak\vspace*{-1.1\baselineskip}
}
\end{minipage}
\n{\footnotesize\baselineskip=10pt 
elliptical and spiral galaxies,  respectively (Odewahn \etal\ 1996). 
Orange and black squares indicate hierarchical size simulations (Kawata \etal
2003). Note that most galaxies at \JAB\cge 28 mag are expected to be smaller
than the HST and JWST diffraction limits (\ie r$_{hl}$\cle 0\arcspt 1).
} 

\noindent galaxies at \JAB\cge 28 mag are
likely unresolved point sources at r$_{hl}$\cle 0\arcspt 1 FWHM, as suggested
by hierarchical size simulations in Fig. 3 (Kawata \etal\ 2002). This is why
they are best imaged from space, which provides the best point-source and
SB-sensitivity in the near-IR. The fact that many faint objects remain
unresolved at the HST diffraction limit effectively reduces the (1+z)$^{4}$
SB-dimming to a (1+z)$^{2}$ flux-dimming (with potentially an intermediate case
for partially resolved objects, or linear objects that are resolved in only one
direction), mitigating the incompleteness of faint galaxy samples. The trick in
deep HST surveys is therefore to show that this argument has not become
circular, and that larger galaxies at high redshift are not missed. Other
aspects that compound these issues are size-overestimation due to object
confusion, size-bias due to the sky background and due to image noise, which
will be studied in detail elsewhere  (\eg Hathi \etal\ 2007). 


\section{What has been done with the Hubble Space Telescope? }

One of the remarkable discoveries by HST was that the numerous faint blue
galaxies are in majority late-type (Abraham \etal\ 1996, Glazebrook \etal\
1995, Driver \etal\ 1995) and small (Odewahn \etal\ 1996, Pascarelle \etal\
1996) star-forming objects. They are the building blocks of the giant galaxies
seen today. By measuring their distribution over rest-frame type versus
redshift, HST has shown that galaxies of all Hubble types formed over a wide
range of cosmic time, but with a notable transition around redshifts
z$\simeq$0.5--1.0 (Driver \etal\ 1998, Elmegreen \etal\ 2007). This was done
through HST programs like the Medium-Deep Survey (Griffiths \etal\ 1994), GOODS
(Giavalisco \etal\ 2004), GEMS (Rix \etal\ 2005), and COSMOS (Scoville \etal\
2007). Subgalactic units rapidly merged from the end of reionization to grow
bigger units at lower redshifts (Pascarelle \etal\ 1996). Merger products start
to settle as galaxies  with giant bulges or large disks around redshifts
z$\simeq$1 (Lilly \etal\ 1998, 2007). These evolved mostly passively since
then, resulting in giant galaxies today, possibly because the epoch-dependent
merger rate was tempered at z\cle 1 by the extra expansion induced by $\Lambda$
(Cohen \etal\ 2003). To avoid caveats from the morphological K-correction
(Giavalisco \etal\ 1996, Windhorst \etal\ 2002), galaxy structural
classification needs to done at rest-frame wavelengths longwards of the Balmer
break at high redshifts (Taylor-Mager \etal\ 2007). JWST will make such studies
possible with 0\arcspt 1--0\arcspt 2 FWHM resolution at observed near--IR
wavelengths (1--5 \mum), corresponding to the restframe optical--near-IR at the
median redshift of faint galaxies (\zmed$\simeq$1--2; Mobasher \etal\ 2007). 


\section{First Light, Reionization \& Galaxy Assembly with JWST }\label{s1}

The James Webb Space Telescope (JWST) is designed as a deployable 6.5 meter
segmented IR telescope for imaging and spectroscopy from 0.6 \mum\ to 28 \mum.
After its planned 2013 launch (Mather \& Stockman 2000), JWST will be
automatically deployed and inserted into an L2 halo orbit. It has a nested
array of sun-shields to keep its temperature at \cle 40 K, allowing faint
imaging to AB\cle 31.5 mag ($\simeq$1 nJy) and spectroscopy to AB\cle 29 mag in
the near--mid-IR. Further details on JWST are given by M. Clampin (this
Volume). 

{\bf First Light:}\ The WMAP polarization results imply that the Dark Ages
which started at recombination (z$\simeq$1089) lasted until the First Light
objects started shining at z\cle 20, and that the universe was first reionized
at redshifts as early as z$\simeq$11--17 (Spergel \etal\ 2003; 2006). The epoch
of First Light is thought to have started with Population III stars of
200-300\Msun\ at z\cge 10--20 (Bromm \etal\ 2003). Groupings of Pop III stars 
and possibly their extremely luminous supernovae should be visible to JWST at
z$\simeq$10--20 (Gardner \etal\ 2006).

\ve 

\null\vspace*{-2.0cm}
\n\makebox[\txw]{
\psfig{file=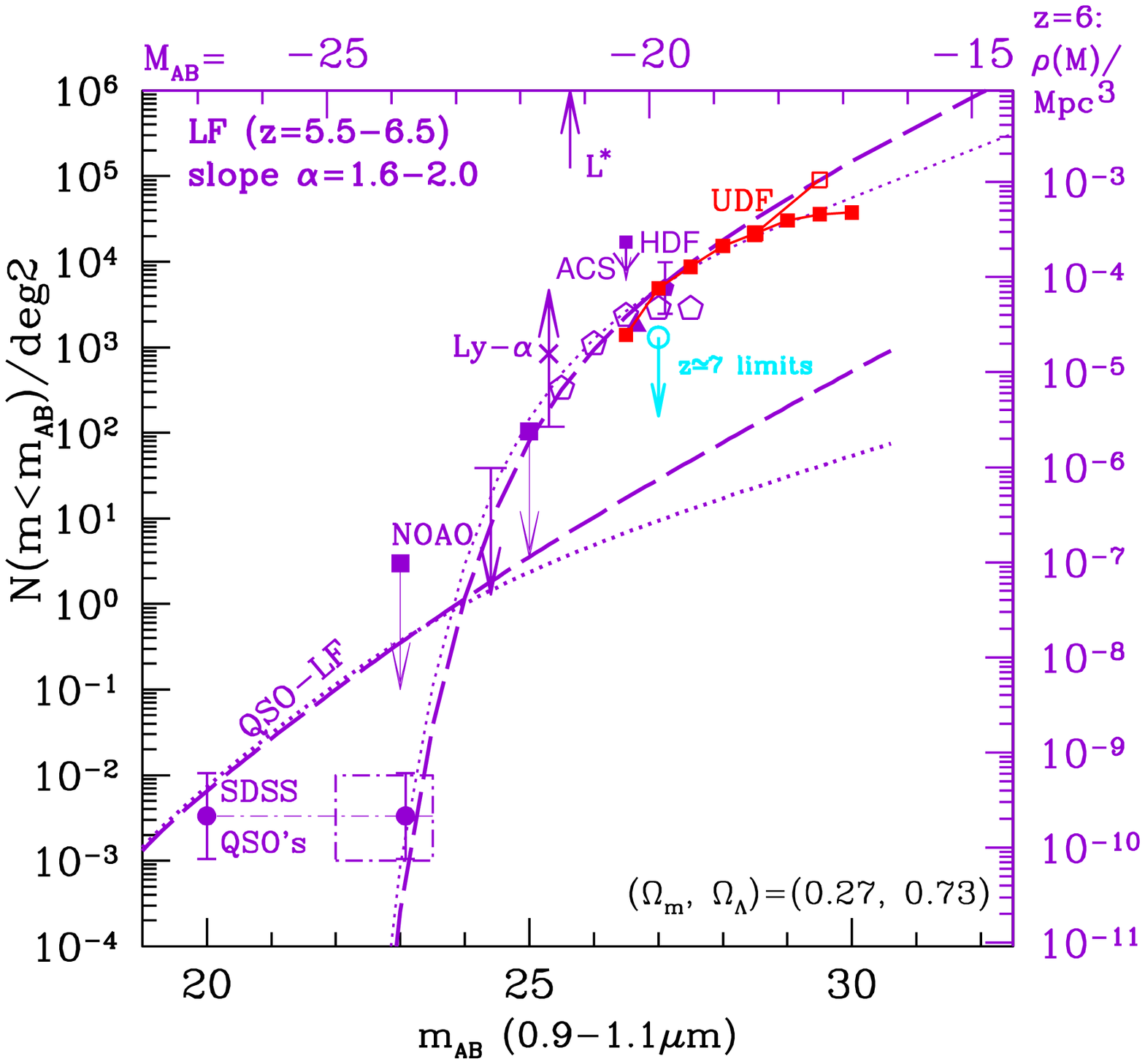,width=0.50\txw}\ \
\psfig{file=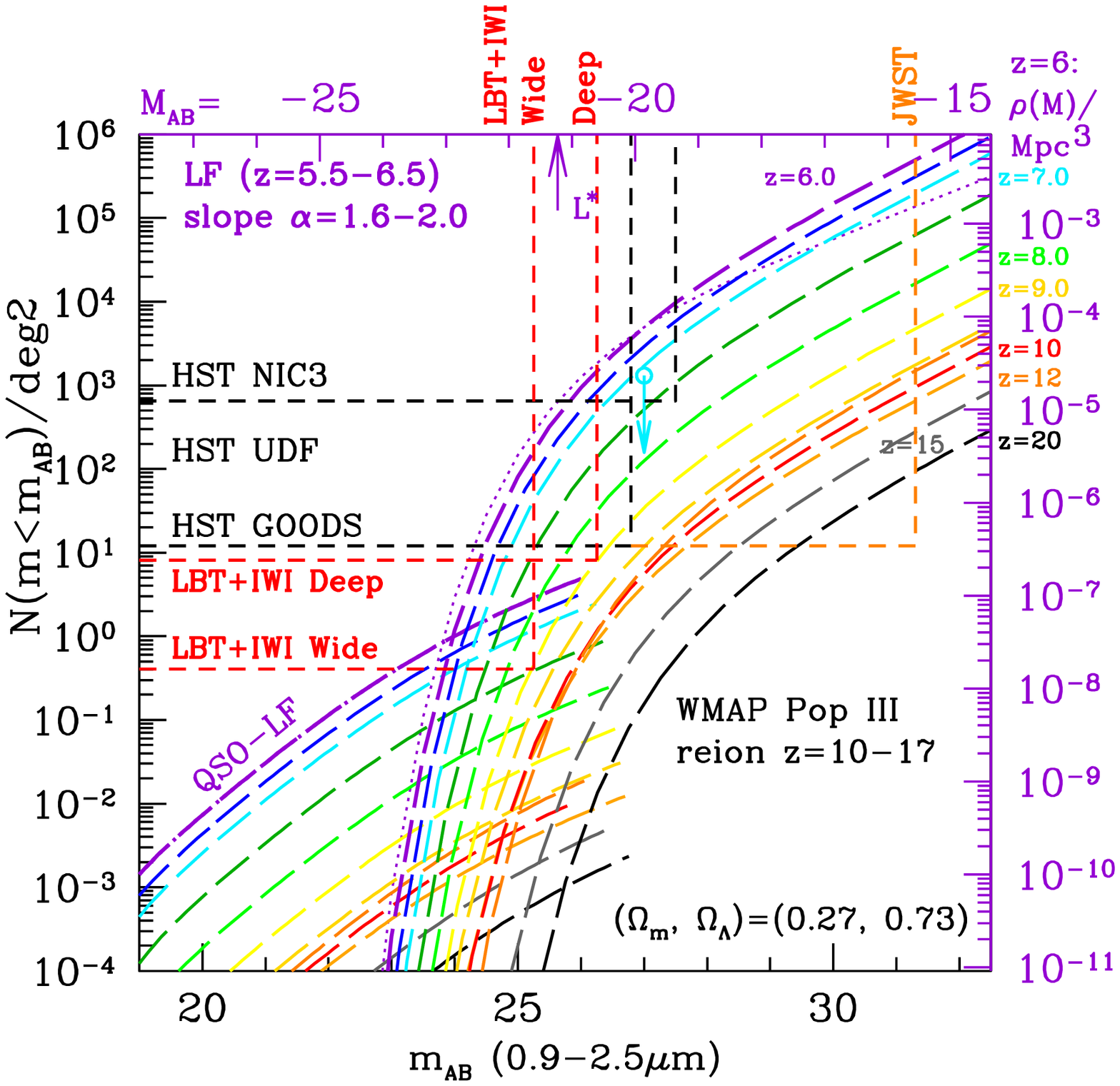,width=0.50\txw}
}
\null
{\footnotesize
\noindent{\bf Fig.~4a} \ Integral luminosity function (LF) of z$\simeq$6 objects,
plotted as surface density vs. AB-mag. The z$\simeq$6 LF may be very steep,
with faint-end Schechter slope $\vert$$\alpha$$\vert$$\simeq$1.8--1.9 (Yan
\& Windhorst 2004b). Dwarf galaxies and not quasars therefore likely
completed the reionization epoch at z$\simeq$6 (Yan \etal\ 2004a). This is
what JWST will observe in detail to AB$\simeq$31.5~mag (1~nJy). \ 
{\bf Fig.~4b} \ Possible extrapolation of the LF of Fig.~4a for z\cge 7,
which is not yet constrained by data. Successive colors show redshift shells
0.5 in $\Delta$z apart from z=6, 6.5, ..., 10, and also for z=12, 15, 20. The
HST/ACS has detected objects at z\cle 6.5, but its discovery space
A$\cdot$$\Omega$$\cdot$$\Delta$log($\lambda$) is limited to z\cle 6.5. NICMOS
similarly is limited to z\cle 8 (Bouwens \etal\ 2004, Yan \& Windhorst 2004b).
JWST can trace the entire reionization epoch from First Light at z$\simeq$20 to
the end of reionization at z$\simeq$6.}

\n This is why JWST needs NIRCam at 0.6--5 \mum\ and MIRI at 5--28 \mum. The
First Light epoch and its embedded Pop III reionizing sources may have been
followed by a delayed epoch of Pop II star-formation, since Pop III supernovae
may have heated the IGM enough that it could not cool and form the IMF of the
first Pop II stars until z\cle 8--10 (Cen 2003). The IMF of Pop II stars may
have formed in dwarf galaxies with masses of 10$^6$--10$^9$\Msun\ with a
gradual onset between z$\simeq$9 and z$\simeq$6. The reionization history may
have been more complex and/or heterogeneous, with some Pop II stars forming in
sites of sufficient density immediately following their Pop III predecessors at
z\cge 10. 

HST/ACS can detect objects at z\cle 6.5, but its discovery space
A$\cdot$$\Omega$$\cdot$$\Delta$log($\lambda$)
cannot trace the entire reionization epoch.
HST/NICMOS similarly is limited to z\cle 8 and provides limited statistics.
HST/WFC3 can explore the redshift range z$\simeq$7--8 with a wider FOV than
NICMOS. Fig. 4b shows that with proper survey strategy (area {\it and} depth),
JWST can trace the LF throughout the entire reionization epoch, starting with
the first star-forming objects in the  First Light epoch at z\cle 20, to the
first star-forming dwarf galaxies at the end of the reionization epoch at
z$\simeq$6. Since in WMAP cosmology the amount of available volume per unit
redshift decreases for z\cge 2, the observed surface density of objects at
z$\simeq$10--20 will be small, depending on the hierarchical model used. This
is illustrated in Fig. 4b, where the predicted surface densities at
z$\simeq$7--20 are uncertain by at least 0.5 dex. To observe the LF of First
Light star-clusters and subsequent dwarf-galaxy formation may require JWST to
survey GOODS-sized areas to AB$\simeq$31.5 mag ($\simeq$ 1 nJy at 10-$\sigma$),
using 7 filters for reliable photometric redshifts, since objects with AB\cge
29 mag will be too faint for spectroscopy. Hence, JWST needs to have the quoted
sensitivity/aperture (``A''; to reach AB\cge 31 mag), field-of-view
(FOV=$\Omega$; to cover GOODS-sized areas), and wavelength range (0.7--28 \mum;
to cover SED's from the Lyman to Balmer breaks at z\cge 6--20), as summarized
in Fig. 4b. 

\n {\bf Reionization:}\ The HUDF data showed that the LF of z$\simeq$6 objects
is potentially very steep (Bouwens \etal\ 2006, Yan \& Windhorst 2004b), with a
faint-end Schechter slope $\vert$$\alpha$$\vert$$\simeq$1.8--1.9 after
correcting for sample incompleteness (Fig. 4a). Deep HST/ACS grism spectra
confirmed that 85--93\% of HUDF i-band dropouts to \zAB\cle 27 mag are at
z$\simeq$6 (Malhotra \etal\ 2005). The steep faint-end slope of the z$\simeq$6
LF implies that dwarf galaxies may have collectively provided enough UV-photons
to complete reionization at z$\simeq$6 (Yan \& Windhorst 2004a). This assumes
that the Lyman continuum escape fraction at z$\simeq$6 is as large as observed
for Lyman Break Galaxies at z$\simeq$3 (Steidel \etal\ 1999), which is
reasonable --- although not proven --- given the expected lower dust content in
dwarf galaxies at z$\simeq$6. Hence, dwarf galaxies, and not quasars, likely
completed the reionization epoch at z$\simeq$6. The Pop II stars in dwarf
galaxies therefore cannot have started shining {\it pervasively} much before
z$\simeq$7--8, or no neutral H-I would be seen in the foreground of z\cge 6
quasars (Fan \etal\ 2003), and so dwarf galaxies may have ramped up their
formation fairly quickly from z$\simeq$9 to z$\simeq$6. A first glimpse of this
may already be visible in the HUDF NICMOS surveys, which suggests a
significantly {\it lower} surface density of z\cge 7 candidates compared to
z$\simeq$6 objects (Bouwens \etal\ 2004; Yan \etal\ 2004b; light blue upper
limit in Fig. 4ab), although the \cge 600 HST orbits spent on the HUDF only
resulted in a few believable z\cge 7 candidates at best. JWST surveys are
designed to provide \cge 10$^4$ objects at z$\simeq$7 and 100's of objects in
the epoch of First Light and at the start of reionization (Fig. 4b).

\n {\bf Galaxy Assembly:}\ JWST can measure how galaxies of all types formed
over a wide range of cosmic time, by accurately measuring their distribution
over rest-frame optical type and structure as a function of redshift or cosmic
epoch. HST/ACS has made significant progress at z$\simeq$6, surveying very
large areas (GOODS, GEMS, COSMOS), or using very long integrations (HUDF,
Beckwith \etal\ 2006). Fourier Decomposition (FD) is a robust way to measure
galaxy morphology and structure in a quantitative way (Odewahn \etal\ 2002),
where even Fourier components indicate symmetric parts (arms, bars, rings), and
odd Fourier components indicate asymmetric parts (tidal features, spurs,
lopsidedness, etc.). FD of nearby galaxies imaged with HST in the rest-frame UV
(Windhorst \etal\ 2002) can be used to quantitatively measure the presence and
evolution of bars, rings, spiral arms, and other structural features at higher
redshifts (e.g., Jogee et al. 2004), and can be correlated to other
classification parameters, such as CAS (Conselice 2003). Such techniques will
allow JWST\linebreak


\null\vspace*{-1.0cm}
\n\makebox[\txw]{
\psfig{file=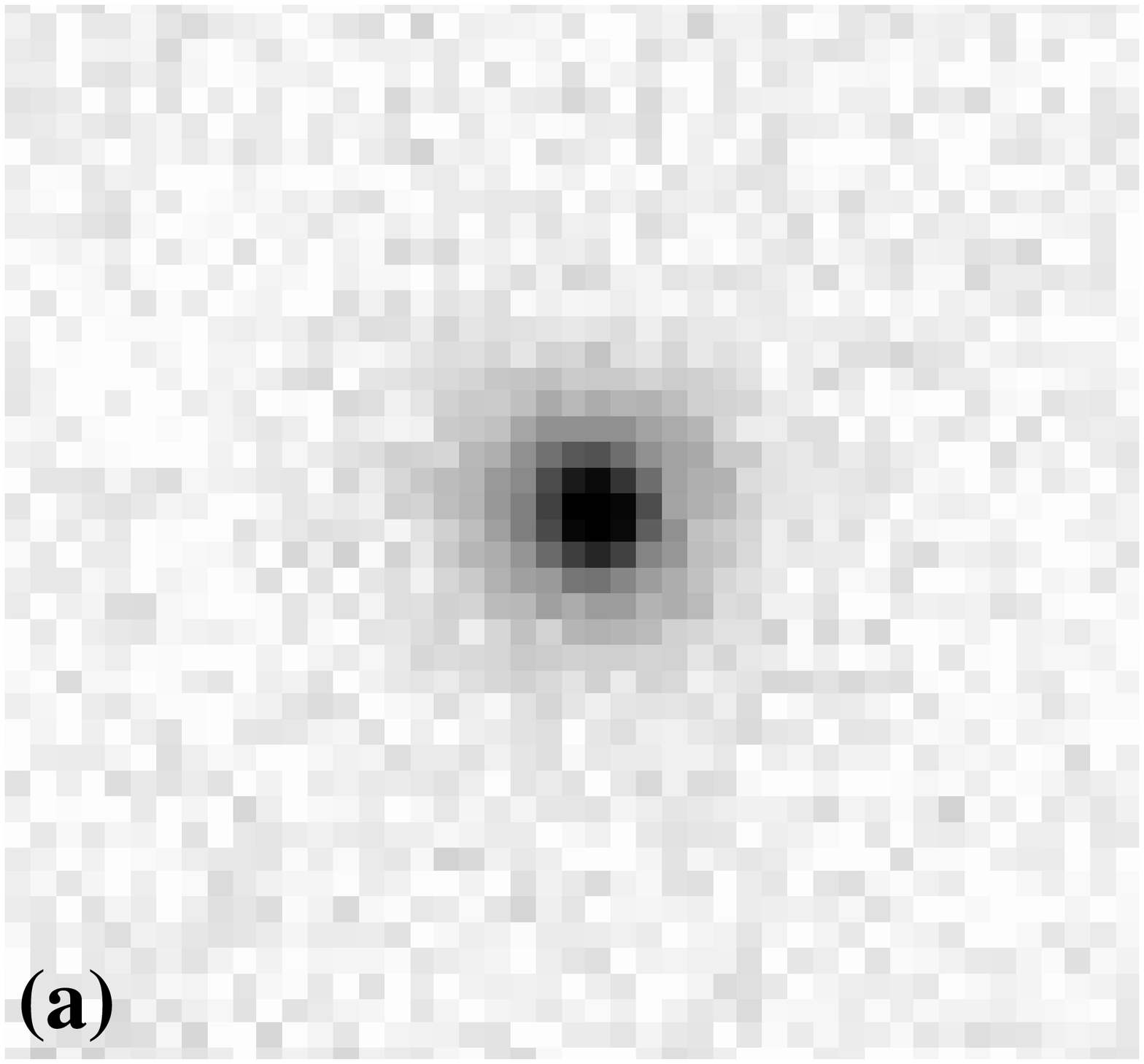,height=0.24\txh}\hfill
\psfig{file=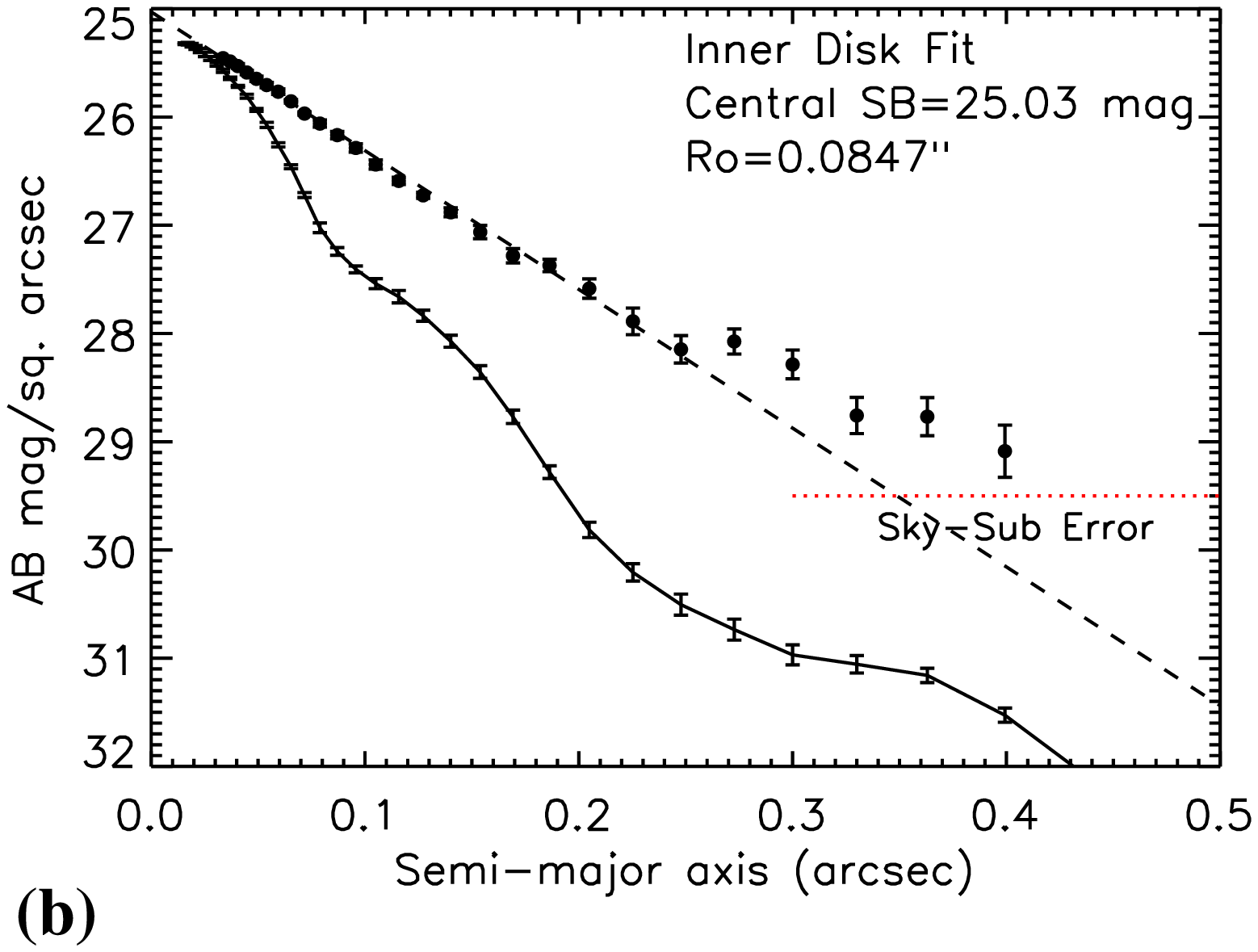,height=0.24\txh}
}
{\footnotesize
{\bf Fig.~5a} \ Sum of 49 compact isolated i-band dropouts in the HUDF,
selected by Hathi \etal\ (2007) from the list of Yan \& Windhorst (2004b).
This image is equivalent to a 5000 hr HST $z$-band exposure --- or a 330\,hr
JWST 1\,\mum\ exposure --- of an average compact isolated z$\simeq$6 object. \ 
{\bf Fig.~5b} \ The radial surface brightness profile of the image stack of
Fig.~5a compared to the ACS PSF. The physical radius where the profile
starts to deviate from a pure exponential profile (dashed) constrains the
dynamical age to $\tau_{dyn}$$\simeq$100--200~Myr at z$\simeq$6, i.e.,
similar to the SED age.}

\n to measure the detailed history of galaxy assembly in the epoch
z$\simeq$1--3, when most of today's giant galaxies were made. JWST will be able
to do this out to z$\simeq$10-15 at least (see Fig. 6 of Windhorst \etal\
2006), hence enabling to quantitatively trace galaxy assembly. The rest-frame
UV-morphology of galaxies is dominated by young and hot stars, as modulated by
copious amounts of intermixed dust. This complicates the study of very high
redshift galaxies. At longer wavelengths (2--28 \mum), JWST will be able to map
the effects from dust in star-forming objects at high redshifts. 

Fig. 5a shows the sum of 49 compact isolated i-band dropouts in the HUDF (Yan
\& Windhorst 2004b), which is a stack of about half the z$\simeq$6 objects that
have no obvious interactions or neighbors. These objects all have similar
fluxes and half-light radii (\re), so this image represents a 5000 hr HST/ACS
z-band exposure-stack on an ``average compact isolated z$\simeq$6 object'',
which is equivalent to a $\sim$330 hr JWST 1 \mum\ exposure on {\it one} such
object. Fig. 5b suggests that the radial SB-profile of this stacked image
deviates from a pure  exponential profile for r\cge 0\arcspt 25, at SB-levels
that are well above those corresponding to PSF and sky-subtraction errors. In
hierarchical models, this physical scale-length may constrain the dynamical age
of these compact isolated i-band z$\simeq$6 dropouts, suggesting that
$\tau_{dyn}$ $\simeq$100-200 Myr for the typical galaxy masses seen at \zAB\cle
29 mag. This age is similar to stellar population age, as discussed in Hathi
\etal\ (2007). This then suggests that the bulk of their stars observed at
z$\simeq$6 may have started forming at z\cle 7--8. This is consistent with the
double reionization model of Cen (2003), where the first reionization by 
Pop\,III stars at z$\simeq$10--20 is followed by a delayed onset of Pop\,II
star-formation in dwarf galaxies at z\cle 9. 


The red boundaries in Fig. 4b indicate part of the galaxy and QSO LF that a
ground-based 8m class telescope with a wide-field IR-camera can explore to
z\cle 9 and AB\cle 25 mag. A ground-based {\it wide-field} near-IR survey to
AB\cle 25--26 mag can sample L$>>$\Lstar galaxies at z\cle 9, which is an
essential ingredient to study the co-evolution of supermassive black-holes and
proto-bulges for z\cle 9, and an essential complement to the JWST First Light
studies. The next generation of wide-field near-IR cameras on ground-based 
8--10 m class telescopes can do such surveys over many \degsq\ to
AB$\simeq$25--26 mag, complementing JWST, which will survey GOODS-sized areas
to AB\cle 31.5 mag (Fig. 4b). 

\section{Conclusions }\label{s2}

High resolution imaging of high redshift galaxies is best done from space,
because faint galaxies are small (\rhl\cle 0\arcspt 15), while the
ground-based sky is too bright and the PSF not stable enough to obtain good
high-resolution images at faint fluxes (AB\cge 27 mag). Ground-based AO imaging
can provide higher spatial resolution on brighter objects than space-based
imaging. HST has led the study of galaxy assembly, showing that galaxies form
hierarchically over time through repeated mergers with sizes growing steadily
over time as r$_{\rm hl}$(z) $\propto$ r$_{\rm hl}$(0)$\cdot$(1+z)$^{\rm -s}$
and s$\simeq$ 1. The Hubble sequence thus gradually emerges at z\cle 1--2, when
the epoch-dependent merger rate starts to wind down. The global onset of Pop
II-star dominated dwarf galaxies ended the process of reionization at
z$\simeq$6. JWST will extend these studies into the epoch of reionization and
First Light, and trace galaxy SED's in the restframe-optical for z\cle 20. In
conclusion, high resolution imaging of high redshift galaxies has made
significant steps forward with HST and recent ground-based AO facilities, and
will see tremendous breakthroughs with JWST and MCAO in the future.

\n This work was supported by HST grants from STScI, which is operated by AURA
for NASA under contract NAS 5-26555, and by NASA JWST grant NAG 5-12460. Other
JWST studies are at:\ www.asu.edu/clas/hst/www/jwst/. We thank Harry Ferguson, 
Jason Melbourne, and Eric Steinbring for helpful suggestions. 









\hspace*{-0.0cm}\vspace*{-1.0cm}
\n 
\eject

\baselineskip=12pt

\end{document}